\documentclass[modern, letterpaper]{aastex62}

\usepackage{amsmath}
\usepackage{amsfonts}
\usepackage{amssymb}
\usepackage{booktabs}

\usepackage{graphicx}
\usepackage{color}

\definecolor{cbblue}{HTML}{3182bd}
\usepackage{hyperref}
\definecolor{linkcolor}{rgb}{0.02,0.35,0.55}
\definecolor{citecolor}{rgb}{0.45,0.45,0.45}
\hypersetup{colorlinks=true,linkcolor=linkcolor,citecolor=citecolor,
            filecolor=linkcolor,urlcolor=linkcolor}
\hypersetup{pageanchor=true}

\newcommand{\sectionname}{Section}
\renewcommand{\figurename}{Figure}
\newcommand{\eqname}{Equation}
\renewcommand{\tablename}{Table}

\newcommand{\package}[1]{\textsl{#1}}
\newcommand{\acronym}[1]{{\small{#1}}}

\newcommand{\project}[1]{\textsl{#1}}

\newcommand{\dd}{\mathrm{d}}

\newcommand{\msun}{\ensuremath{\mathrm{M}_\odot}}

\definecolor{mahogany}{RGB}{165,15,21}

\graphicspath{{}}

\newcommand{\apogee}{\project{\acronym{APOGEE}}}
\newcommand{\sdssiv}{\project{\acronym{SDSS-IV}}}

\newcommand{\DR}{\acronym{DR14}}

\newcommand{\logg}{\ensuremath{\log g}}
\newcommand{\Teff}{\ensuremath{T_{\textrm{eff}}}}
\newcommand{\med}[1]{\ensuremath{\textrm{med}\left(#1\right)}}
\newcommand{\Psurf}{\ensuremath{P_\textrm{surface}}}

\newcommand{\nunimodal}{320}
\newcommand{\nclean}{234}

\shortauthors{Price-Whelan et al.}

\begin{document}\sloppy\sloppypar\raggedbottom\frenchspacing 

\title{Binary companions of evolved stars in \apogee\ \DR: \\
       Orbital circularization}

\author[0000-0003-0872-7098]{Adrian~M.~Price-Whelan}
\affiliation{Department of Astrophysical Sciences,
             Princeton University, Princeton, NJ 08544, USA}
\email{adrn@astro.princeton.edu}
\correspondingauthor{Adrian M. Price-Whelan}

\author[0000-0002-6710-7748]{Jeremy~Goodman}
\affiliation{Department of Astrophysical Sciences,
             Princeton University, Princeton, NJ 08544, USA}



\begin{abstract}\noindent 
Short-period binary star systems dissipate orbital energy through tidal
interactions that lead to tighter, more circular orbits.
When at least one star in a binary has evolved off of the main sequence, orbital
circularization occurs for longer-period ($\approx 100~\textrm{day}$) systems.
Past work by \citet{Verbunt:1995} has shown that the orbital parameters and the
circularization periods of a small sample of binary stars with evolved-star
members can be understood within the context of standard tidal circularization
theory.
Using a sample of binaries with subgiant, giant, and red clump star members that
is nearly an order of magnitude larger, we reexamine predictions for tidal
circularization of binary stars with evolved members.
We confirm that systems predicted by equilibrium-tide theory to have circular
orbits generally have negligible measured eccentricities.
The circularization period is correlated with the surface gravity (i.e. size) of
the evolved member, indicating that the circularization timescale must be
shorter than the evolutionary timescale along the giant branch.
A few exceptions to the conclusions above are mentioned in the discussion: Some
of these exceptions are likely systems in which the spectrum of the secondary
biases the radial velocity measurements, but four appear to be genuine,
short-period, moderate-eccentricity systems.

\end{abstract}

\keywords{
  binaries:~spectroscopic
  --
  binaries:~close
  --
  stars:~evolution
  --
  stars:~interiors
}

\section{Introduction} \label{sec:intro}

From studies of binary star systems in open clusters, it is clear that
short-period binaries tend to have smaller eccentricities as compared to
longer-period binaries of the same age (e.g., \citealt{Mathieu:2005}).
Within a given population of binaries, this manifests as an apparently steep
transition from a spread in eccentricities to mainly circularized orbits.
The transition occurs around a characteristic period that depends on the age and
evolutionary state of the population (see, e.g., \figurename~5 in
\citealt{Mathieu:2005}).
Many studies have measured the ``circularization period,'' predominantly for
main-sequence binaries (e.g., \citealt{Latham:2002, Meibom:2006,
Kjurkchieva:2017}), and have found that it tends to be between 5--20 days,
depending on age (e.g., \citealt{Mathieu:1988}).
This also appears to be consistent with the period and eccentricity
distributions for ``heartbeat'' stars (\citealt{Shporer:2016}).
For giant stars, the circularization period appears to be longer, closer to
$\approx 100$ days (e.g., \citealt{Mayor:1984, Bluhm:2016}), but fewer systems
with giant star members have been studied.
These observed trends are likely a result of orbital circularization rather than
a manifestation of binary star formation, as the circularization or transition
period appears to vary with the age of the population (\citealt{Meibom:2005}).

Theories that explain orbital circularization generally rely on tidal
dissipation to predict the timescale and efficiency of this phenomenon (see
\citealt{Mazeh:2007hp, Zahn:2008} for recent reviews).
For stars with deep convective zones (and especially evolved stars), the
equilibrium tide theory for circularization (\citealt{Zahn:1977, Zahn:1989}) has
been shown to reasonably reproduce the circularization periods of a small sample
of binaries with giant star members in open clusters (\citealt{Verbunt:1995}).
In this theory, the tidal bulge induced on the primary (evolved) star will lag
the orbital motion of the companion because of coupling of the tidal flow to
turbulent eddies driven by convection.
These eddies cause an effective viscosity in the convective region of the
primary star, and the magnitude of this viscosity affects the amount of lag,
which directly relates to the circularization timescale for a given binary
system (\citealt{Zahn:1989}).
This viscous dissipation of orbital energy acts to synchronize and circularize a
binary system, and align the rotational and orbital axes (\citealt{Zahn:1977,
Zahn:1989}).

In the context of equilibrium tide theory (e.g., \citealt{Zahn:1989}), the
time-dependence of the eccentricity, $e$, and semi-major axis, $a$, of a binary
star system is given by
\begin{align}
    \frac{1}{t_\textrm{circ}} &= f \,
        \left(\frac{L_1}{M_{\textrm{env}} \, R_1^2}\right)^{1/3} \,
        \frac{M_{\textrm{env}}}{M_1} \,
        q \, (1 + q) \,
        \left(\frac{R_1}{a}\right)^8 \label{eq:tcirc}\\
    \frac{1}{e} \, \frac{\dd e}{\dd t} &= - \frac{1}{t_\textrm{circ}}
        \label{eq:dlne} \\
    \frac{1}{a} \, \frac{\dd a}{\dd t} &= - \frac{38}{7} \, e^2 \,
        \frac{1}{t_\textrm{circ}} \label{eq:dlna}
\end{align}
where $L_1, M_1, R_1$ are the luminosity, mass, and radius of the primary
(evolved) star, $M_{\textrm{env}}$ is the mass of the convective envelope of the
primary, $q = M_2 / M_1$ is the binary mass ratio, $a$ is the binary semi-major
axis, and $f$ is a dimensionless factor of order unity that depends on the
convective and dissipative properties of the convective envelope
(\citealt{Zahn:1977, Zahn:1989, Verbunt:1995}).
Following \citet{Verbunt:1995}, we set $f=1$.
Note that the expressions above are derived assuming $e \ll 1$:
$t_\textrm{circ}$ is expected to be somewhat shorter for high-eccentricity
systems (e.g., \citealt{Hut:1981}).

The steep $\left(R_1 / a\right)^8$ scaling in \eqname~\ref{eq:tcirc}
implies that, for a given binary, even small changes in the radius of the
primary results in very large changes to the circularization timescale.
For example, for binary stars with periods $P \gtrsim 10~\textrm{day}$, we
expect orbital circularization to occur rapidly when the primary begins to
evolve.
Further, if the circularization timescale (\eqname~\ref{eq:tcirc}) remains short
compared to the evolutionary timescale along the subgiant and giant branches,
then the circularization period of a sample of binary stars should correlate
with the present-day radius or evolutionary state of the primary.

Using a sample of $>200$ binary star systems with at least one evolved member,
we study the circularization period along the giant branch.
We show that indeed the inferred circularization period is a function of the
surface gravity of the primary star, and thus that orbital circularization must
occur faster than post-main-sequence stellar radius and structure evolution.



\section{Data} \label{sec:data}

\begin{figure}[tb]
\begin{center}
\includegraphics[width=\textwidth]{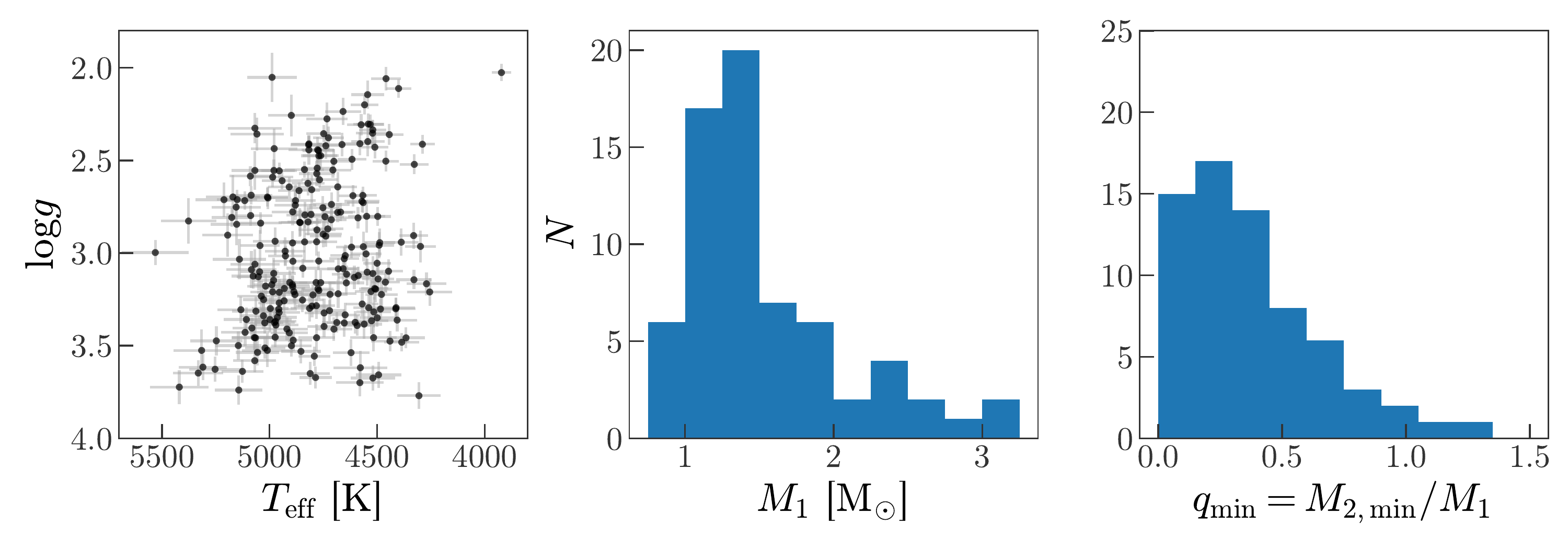}
\end{center}
\caption{%
\textit{Left}: Stellar parameters for all \nclean\ systems used in this work.
Effective temperatures, \Teff, and surface gravities, \logg, are from the
\apogee\ \DR\ catalog.
\textit{Middle}: Primary masses for 68 systems with masses estimated in
\cite{Ness:2015}.
\textit{Right}: Minimum mass ratios computed using primary masses from
\cite{Ness:2015} and companion masses estimated from orbital parameters from
\cite{Price-Whelan:2018}.
\label{fig:apogee}
}
\end{figure}

To identify binaries, we use sources with repeat radial velocity measurements in
data release 14 (\DR) of the \apogee\ survey
(\citealt{Majewski:2017,Abolfathi:2017}), a component of the Sloan Digital Sky
Survey IV (\sdssiv; \citealt{Gunn:2006,Blanton:2017}).
A full description of our search methodology and binary star catalogs can be
found in companion work (\citealt{Price-Whelan:2018}).

Briefly, we use a custom-built Monte Carlo sampler (\package{The Joker};
\citealt{Price-Whelan:2017}) to generate posterior samples over binary orbital
parameters (period, eccentricity, etc.) for all stars with $\geq 3$ radial
velocity measurements in \apogee\ \DR\ that pass a series of quality cuts.
By making cuts on our posterior belief about the amplitude of radial velocity
variations, we identified $\sim$5,000 binary star systems with at least one
evolved member.
However, the majority of these systems have too few radial velocity measurements
to uniquely determine the binary orbital parameters.

Here, we consider only \nunimodal\ binary systems for which the period and
eccentricity can be uniquely determined (the ``high-$K$, unimodal'' sample of
\citealt{Price-Whelan:2018}).
We further subselect the \nclean\ primary stars with $\logg > 2$ that pass
visual inspection (from previous work, \texttt{clean\_flag == 0}; see
\sectionname~5.2 in \citealt{Price-Whelan:2018}).
We note that because of the sparse time sampling of the \apogee\ survey, we
expect that our detection efficiency for high-eccentricity systems is poor, but
do not expect biases for low-eccentricity systems.
\figurename~\ref{fig:apogee}, left panel, shows the stellar parameters for all
\nclean\ primary stars in the sample used in this work, middle panel shows the
distribution of primary masses for the 68 systems with masses from
\cite{Ness:2015}, and right panel shows the minimum mass ratios for the 68
systems.

\section{Orbital circularization of \apogee\ binaries}
\label{sec:results}

\begin{figure}[t]
\begin{center}
\includegraphics[width=\textwidth]{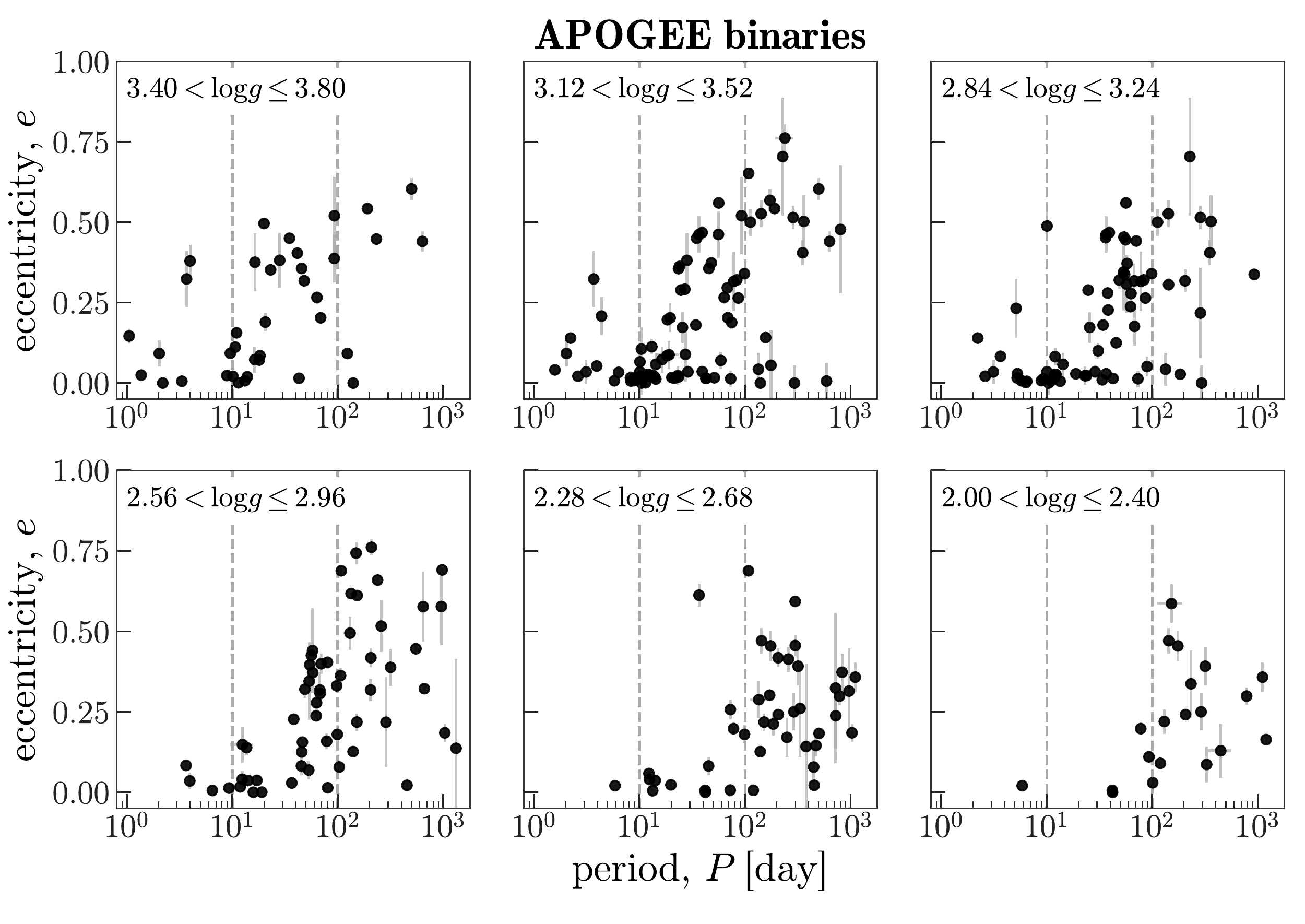}
\end{center}
\caption{%
Orbital period and eccentricity for all systems considered in this work.
Each panel shows systems in the specified bin of primary surface gravity, \logg.
Vertical gray lines show periods of $10~\textrm{day}$ and $100~\textrm{day}$:
binaries with subgiant members (top left) show circularization periods close to
$10~\textrm{day}$, whereas binaries with RGB members (bottom right) show
circularization periods close to $100~\textrm{day}$.
Intermediate bins show steadily increasing circularization periods with
decreasing \logg.
\label{fig:P-e-grid}
}
\end{figure}

Our sample of binaries contains primary stars with a range of stellar parameters
(\figurename~\ref{fig:apogee}), and therefore a range of expected
circularization periods.
\figurename~\ref{fig:P-e-grid} shows orbital period and eccentricity for all
systems in bins of primary surface gravity:
From top left to bottom right shows bins of decreasing surface gravity, i.e.
from subgiants to giant branch stars.
Vertical dashed lines at $10~\textrm{day}$ and $100~\textrm{day}$ are meant as
reference lines.
Note the steadily increasing circularization period from top left to bottom
right as the typical size of the primary increases.

\begin{figure}[t]
\begin{center}
\includegraphics[width=\textwidth]{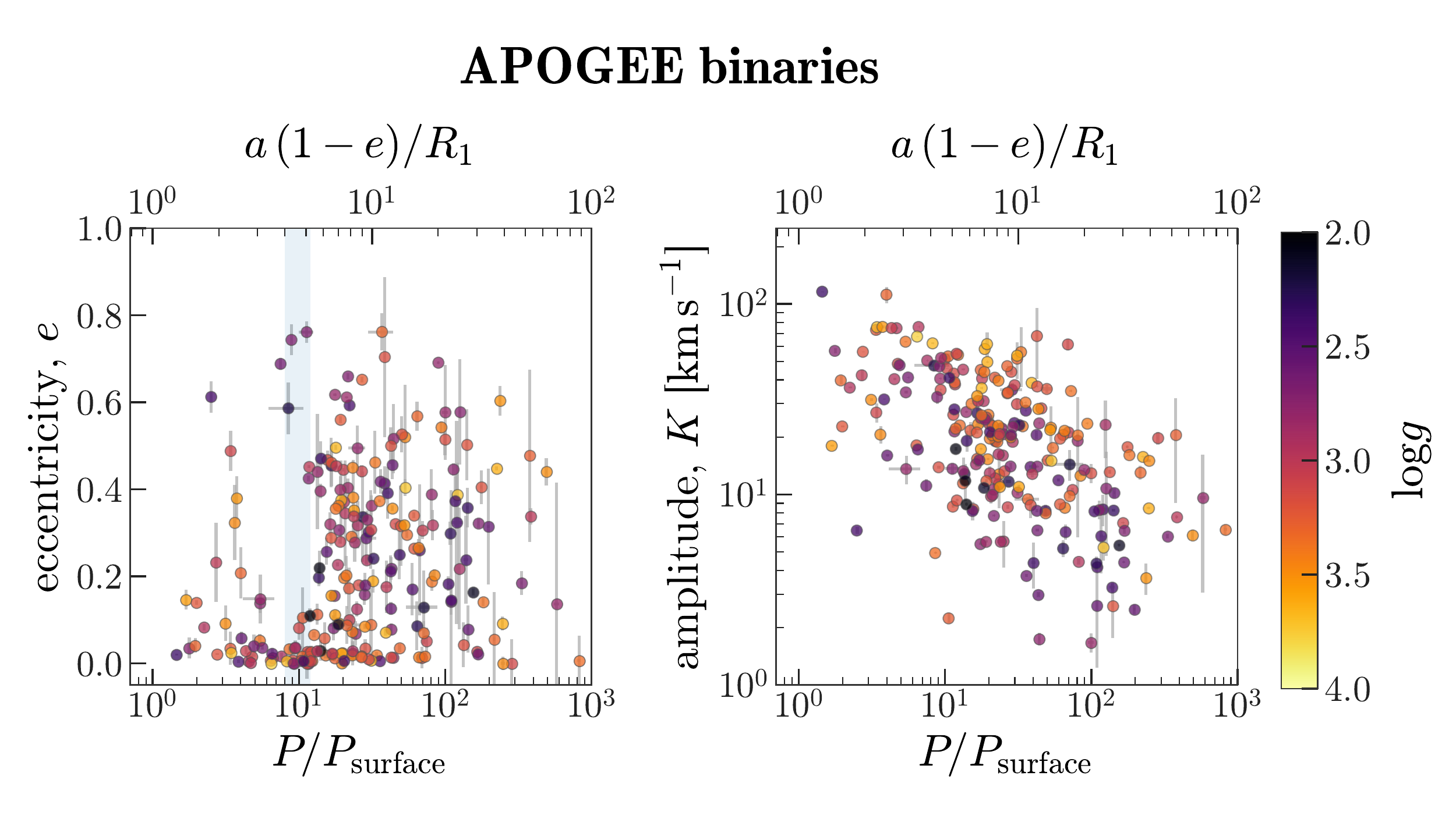}
\end{center}
\caption{%
\textit{Left panel:} Orbital period and eccentricity for binary star systems
with evolved members, with period values normalized by the orbital period
of a hypothetical companion that grazes the surface of each primary star,
$P_{\textrm{surface}}$, assuming $M_1 = 1.36~\msun$ and $M_2 = 0.5~\msun$.
The sharp transition from eccentric systems to almost all circular orbits at
$P/P_\textrm{surface} \approx 10$ is likely an outcome of tidal circularization.
\textit{Right panel:} Normalized period and inferred radial velocity amplitude
for the same systems.
\label{fig:PeK}
}
\end{figure}

To remove the dependence on the size of the primary, \figurename~\ref{fig:PeK}
(left) shows period and eccentricity but now with periods normalized by the
orbital period at which the minimum separation of the binary is equal to the
surface size of the primary star,
\begin{align}
    \Psurf &= 2\pi \,
        \left(\frac{G \, (M_1+M_2)}{R_1^3}\right)^{-1/2} \,
        \left(1-e\right)^{-3/2}
    \quad . \label{eq:Psurf}
\end{align}
We compute \Psurf\ for all systems under the following assumptions:
\begin{description}
    \item[Primary/companion masses] Only a small fraction of the systems in our
    sample have measured primary masses (\figurename~\ref{fig:apogee}) and
    therefore have measured (minimum) mass ratios.
    We make the simplifying assumption that all primary stars have masses equal
    to the median mass over all stars with prior mass measurements in this
    sample, $\med{M_1} = 1.36~\msun$, and all companions have masses equal to
    the median over minimum companion masses, $\med{M_{2, \textrm{min}}} =
    0.5~\msun$.
    \item[Primary evolved first] We assume that the primary star (i.e. the
    observed star) is the first member of the binary to evolve off the main
    sequence.
    \item[First ascent] We assume that all primary stars are on their first
    ascent up the giant branch.
    This is motivated by the small fraction of red clump stars in our sample:
    Only 7 of the primaries were identified as confident red clump stars in a
    recent study of \apogee\ \DR\ evolved star (\citealt{Ting:2018}).
\end{description}

Assuming that the semi-major axes of the \apogee\ binaries have remained
constant, and with the assumptions above, we can also compute the expected
change in eccentricity, $\Delta \ln e$, for each binary (see also \figurename~4
in \citealt{Verbunt:1995}).
As noted in \citet{Verbunt:1995}, the assumption of constant binary semi-major
axis is technically incorrect, but the typical change to the semi-major axis is
only between 1--10\%, i.e. smaller than the uncertainty introduced in the
inferred semi-major axes for our systems due to unknown inclination.

To compute $\Delta \ln e$, we use a model stellar evolution track to integrate
\eqname~\ref{eq:dlne} up to the measured \logg\ of each primary star.
In detail, we use \acronym{MESA} (\citealt{Paxton:2011}) to follow the stellar
evolution of a $M_1 = 1.4~\msun$ star with solar metallicity from the
pre-main-sequence phase to the asymptotic giant branch (AGB) phase (we stop the
models when they first reach $\logg = 0$).
For each \apogee\ star, we use linear interpolation with the output of this
evolutionary model to solve for the eccentricity change starting from the time
the star leaves the main sequence up to the phase at which the model has the
same surface gravity as is measured.
We use 10,000 steps evenly spaced between these two phases and use Simpson's
rule to compute the integral.

\begin{figure}[h]
\begin{center}
\includegraphics[width=\textwidth]{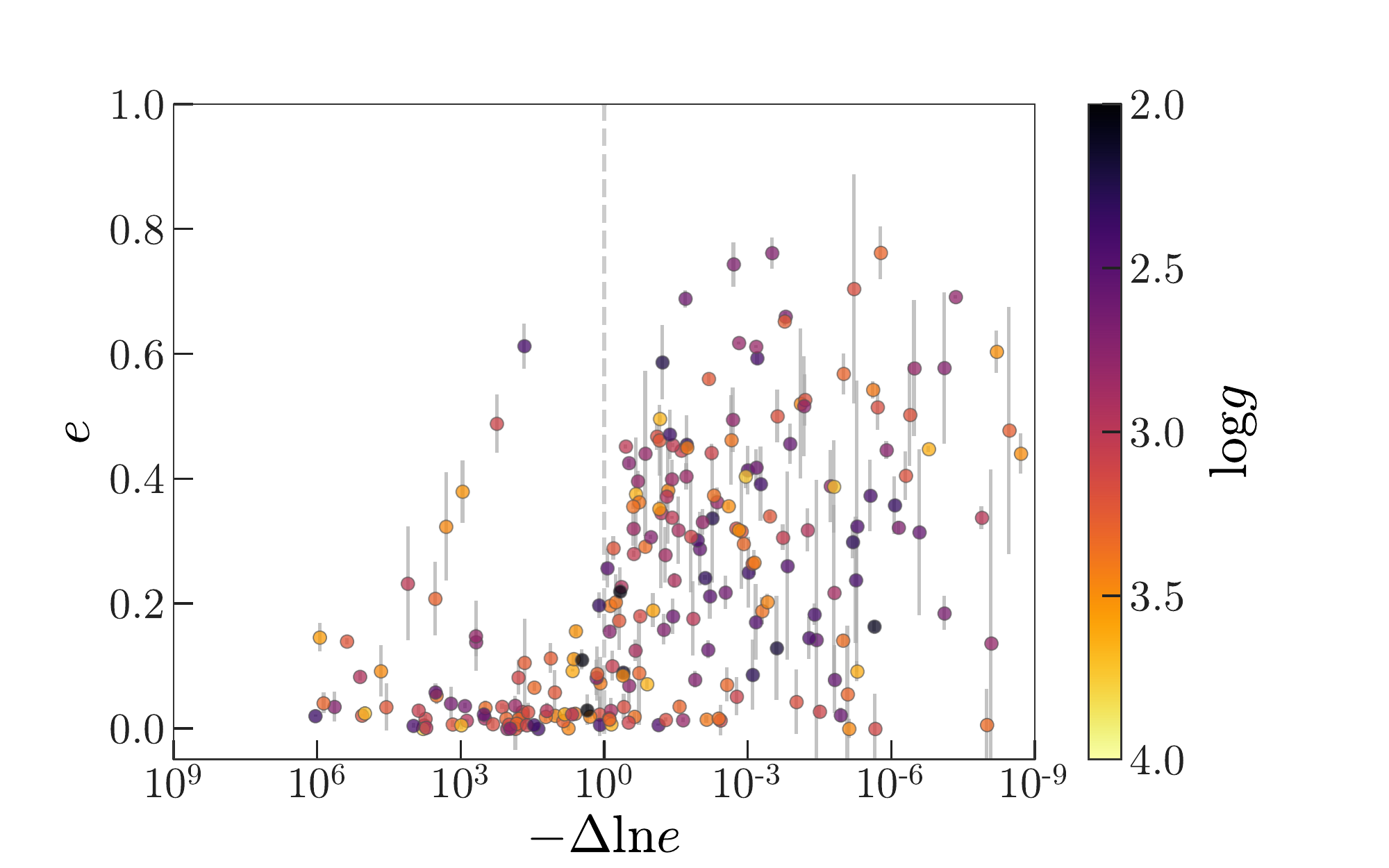}
\end{center}
\caption{%
Predicted change in eccentricity from tidal circularization, $\Delta \ln e$, and
observed eccentricity for all \apogee\ binary systems.
Change in eccentricity is computed assuming that the semi-major axis of the
systems remain constant during eccentricity evolution, and assuming that all
stars have a primary mass of $M1 = 1.4~\msun$ and a companion mass of $M_2 =
0.5~\msun$ (the median values from all systems with measured masses).
Most systems with large predicted circularization have small observed
eccentricities, with the exception of a few outlier systems.
\label{fig:dlne}
}
\end{figure}

\figurename~\ref{fig:dlne} shows the expected change in eccentricity plotted
against the observed eccentricity for all of the \apogee\ binaries.
With the exception of a few outliers, systems that are predicted to have large
negative changes to the log-eccentricity are circular.
This confirms the conclusions of previous work based on a much smaller sample of
giant star binaries (\citealt{Verbunt:1995}): the theory of equilibrium tides
successfully explains the observed eccentricities of close binaries with
member stars that have large convective envelopes.

\section{Population synthesis}
\label{sec:theory}


To compare with the data, we generate a simulated population of binary star
systems with primary stars that have similar stellar parameters (by the end of
their evolution) to our sample of \apogee\ system primaries.
We assume simple initial binary orbital parameter distributions (i.e. in period
and eccentricity) and compute the change in eccentricity and separation of the
companion orbit as the primary star evolves off the main sequence.

In detail, we sample primary stellar masses, companion masses, and primary
surface gravities by fitting two-component Gaussian mixture models (GMMs) to the
stars in our sample with measurements of each, then re-sample using the fitted
distributions.
As described in previous work (\citealt{Price-Whelan:2018}), primary mass
measurements come from \cite{Ness:2015}, secondary (minimum) masses come from
the posterior samplings over orbital parameters using \apogee\ radial velocity
data, and surface gravities come from the \apogee\ (\DR) data reduction pipeline
(\citealt{Garcia-Perez:2016}).
In our sample, the median primary mass, companion mass, and surface gravity are
$\med{M_1} = 1.4~\msun$, $\med{M_{2, \textrm{min}}} = 0.4~\msun$, and
$\med{\logg} = 3.1$.
We assume $M_2 = M_{2, \textrm{min}}$ when generating companion masses.

We generate eccentricities $e$ by sampling from a truncated normal distribution
with mean and standard deviation $(\mu, \sigma) = (0.4, 0.3)$, truncated to the
domain $[0, 1]$.
We note that theoretical predictions of the binary star eccentricity
distribution over this period range would predict a thermalized eccentricity
distribution $p(e) = 2e$ (\citealt{Jeans:1919}), but the observed eccentricity
distributions of main sequence binary star systems with periods $10 < P <
1000~\textrm{day}$ is broadly consistent with being flat for moderate
eccentricities, with fewer very low and high eccentricities
(\citealt{Duchene:2013}).
Our comparison with this simulated population does not depend strongly on this
choice of initial eccentricity distribution.

\begin{figure}[htbp]
\begin{center}
\includegraphics[trim={0 0 1cm 0}, clip, width=\linewidth]{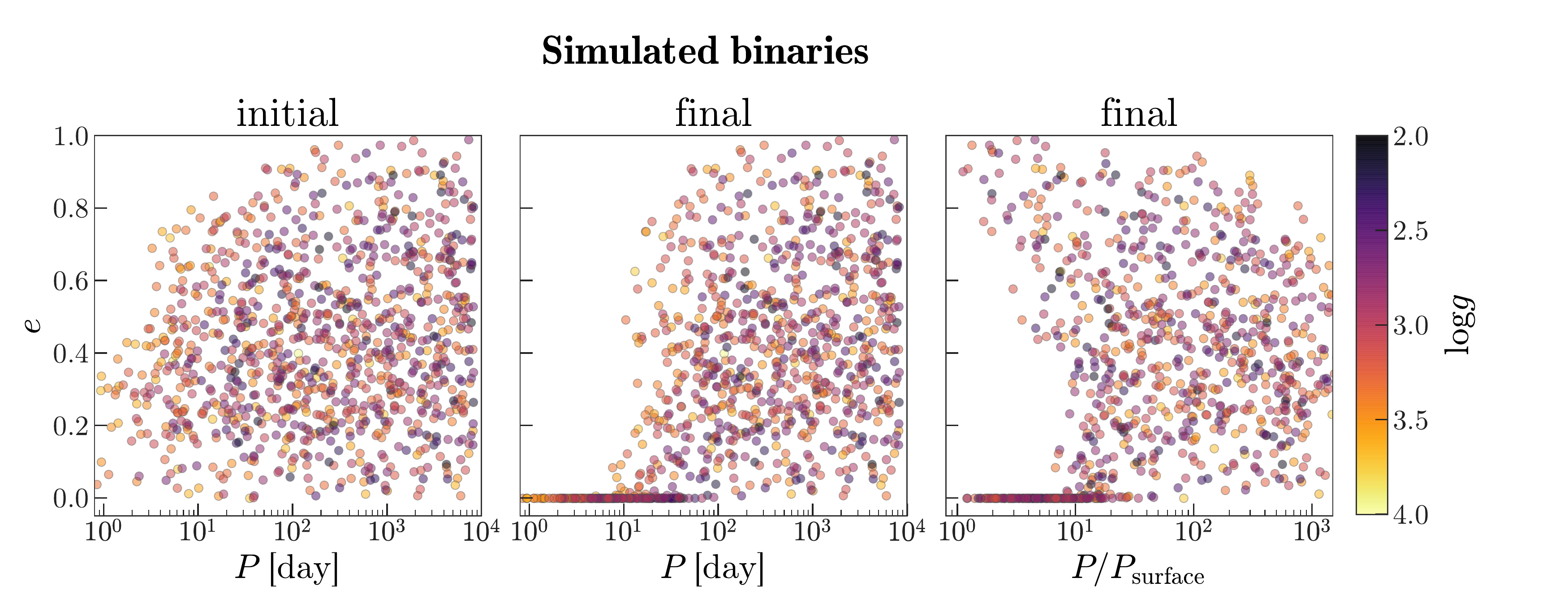}
\end{center}
\caption{%
Simulated binary star systems with one giant star member.
Points are colored by primary mass, $M_1$, and the size indicates the size of
the star, i.e. the surface gravity $\log g$.
\textit{Left:} Distribution of initial orbital parameters period, $P$, and
eccentricity, $e$.
\textit{Middle:} Final distribution of period and eccentricity after computing
the change in eccentricity from evolution of the primary star.
\textit{Right:} Final distribution of period and eccentricity, with period
normalized by \Psurf, the period of a hypothetical, point-mass
companion whose orbit grazes the surface of the primary star at pericenter.
\label{fig:simulated}
}
\end{figure}

We generate initial binary orbital periods, $P$, by assuming a distribution that
is uniform in $\ln P$ between $(\Psurf, 8192)~\textrm{day}$.
\figurename~\ref{fig:simulated} (left) shows the initial periods and
eccentricities of the simulated systems.
Markers are colored by the mass of the primary, $M_1$, and the size of the
marker indicates the log-surface gravity, \logg.

\begin{figure}[h]
\begin{center}
\includegraphics[width=0.5\textwidth]{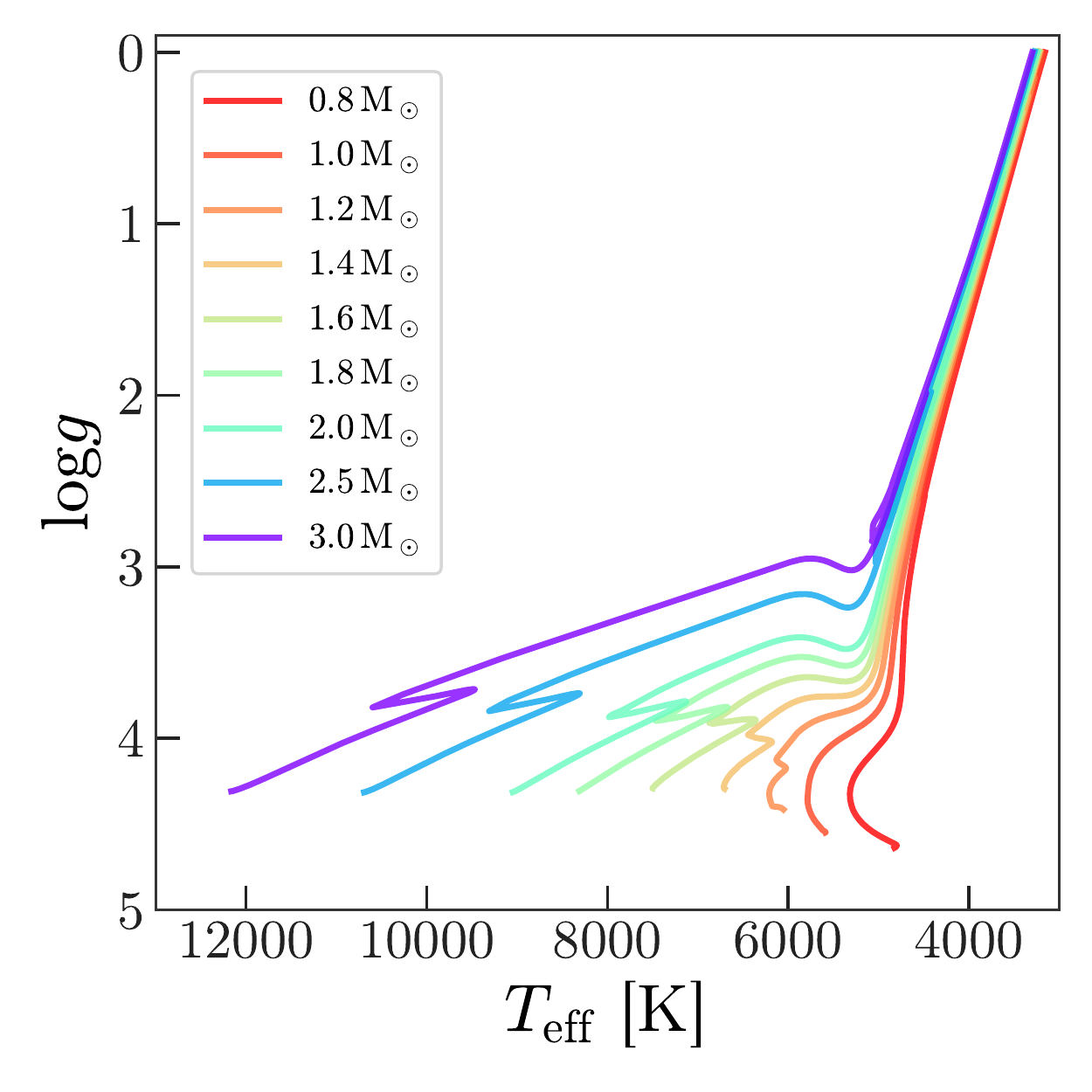}
\end{center}
\caption{%
Evolutionary tracks of the eight stellar models we use to simulate the orbital
evolution of a population of binary star systems with an evolved member.
Horizontal axis shows the effective temperature, $T_\textrm{eff}$, and vertical
axis shows the log-surface-gravity, \logg.
Tracks are colored by the mass of the star.
All models have solar metallicity.
\label{fig:mesa}
}
\end{figure}

To follow the stellar evolution of the primary stars, we run stellar evolution
models using \acronym{MESA} (\citealt{Paxton:2011}) for stars with $M = [0.8, 1,
1.2, 1.4, 1.6, 1.8, 2, 2.5, 3]~\msun$ and solar metallicity.
\figurename~\ref{fig:mesa} shows evolutionary tracks in surface gravity and
effective temperature for each of these models.
We follow the evolution from the pre-main-sequence phase until the AGB phase,
but again only use the post-main-sequence evolution when evolving the orbit of
the binary.
At each timestep during the evolution, we output and store the standard stellar
parameters along with the size and mass of the convective envelope.

We discretize the primary masses generated from the GMMs onto the grid of masses
for which we have \acronym{MESA} models, then use equilibrium tide theory to
compute the change in eccentricity and separation of the companion orbit.
To do this, we solve the coupled differential equations \eqname
s~\ref{eq:dlne}--\ref{eq:dlna} up to the phase of evolution at which the stellar
model has the same (or closest) surface gravity to the simulated primary star.
We use the first-order Euler method with 10,000 time steps between the main
sequence and the given final \logg\ of each primary star, and use linear
interpolation to interpolate the \acronym{MESA} output stellar parameters onto
the integration grid.
We assume that all stars are on their first ascent up the giant branch, which
should underestimate the number of circularized systems with surface gravities
$\logg \sim 2.5$ (i.e. near the red clump, where stars have already reached the
tip of the giant branch).
We assume that the equilibrium tides dominate the circularization process for
all systems and therefore ignore the effect of ``dynamical tides''
(e.g., \citealt{Goodman:1998}) during the main sequence phase.
Finally, we assume that all of our systems are detached binaries.

\figurename~\ref{fig:simulated} (middle) shows the final periods and
eccentricities of the simulated systems.
As expected, the circularization period for higher \logg\ systems (i.e. smaller
radii, lighter markers) appears to be close to $\sim 10~\textrm{day}$, but the
circularization period for stars with lower \logg\ (i.e. larger radii, darker
markers) is closer to $\sim 100~\textrm{day}$.
In the right panel we normalize the orbital period by \Psurf:
This rescaling removes the dependence on primary size or \logg\ and predicts
that circularization should occur around $P / \Psurf \approx 10$.
The simulated systems that remain very eccentric, $e \gtrsim 0.6$, with periods
$P / \Psurf < 10$ are likely a relic of the fact that \eqname
s~\ref{eq:dlne}--\ref{eq:dlna} break down when $e \sim 1$ (e.g.,
\citealt{Hut:1981}).

The sharp transition in eccentricity around $P/\Psurf \approx 10$ or $a/R_1
\approx 4$--5 observed in the sample of \apogee\ binaries
(\figurename~\ref{fig:PeK}, left) is therefore qualitatively consistent with
predictions from this simulated population (\figurename~\ref{fig:simulated},
right).

\section{Discussion} \label{sec:discussion}

\subsection{Assumptions}

Here we return to the assumptions made above and assess their applicability:
\begin{description}
    \item[Primary/companion masses] Most of the systems in our sample of
    binaries have unmeasured primary or companion masses.
    In computing the normalization period, \Psurf, and to estimate the predicted
    change in eccentricity, $\Delta \ln e$, we have thus made simple assumptions
    about the binary component masses (see \emph{Primary/companion masses}
    above).
    It is therefore interesting that we see such a sharp transition in the
    eccentricity distribution (\figurename~\ref{fig:PeK}, left panel), and good
    agreement with predictions from population synthesis and evolution of the
    simulated binaries (\sectionname~\ref{sec:theory}).
    However, this can be understood given the scaling of the circularization
    timescale (\eqname~\ref{eq:tcirc}) and the mass distribution of stars with
    measured masses (\figurename~\ref{fig:apogee})

    Assuming a mass ratio $q = 0.3$, the circularization timescale close to the
    circularization period, $P / \Psurf \approx 10$, is very small compared to
    the lifetime on the giant branch for a $1.4~\msun$ star, $t_\textrm{circ}
    \sim 10^5~\textrm{yr} \ll t_\textrm{RGB} \sim 10^8~\textrm{yr}$.
    The circularization time scales with the inverse mass ratio squared and the
    cubed-root of the primary mass, $t_\textrm{circ} \propto M_1^{1/3} \, \left[
    q \, (1 + q) \right]^{-1}$:
    The small spread in primary masses can therefore be neglected, and
    variations in the mass ratio by typical factors of $\sim 2$
    (\figurename~\ref{fig:apogee}still lead to circularization timescales that
    are much shorter than the giant branch lifetime.
    For the sample of binaries considered here, the simple assumptions about
    primary and companion masses are therefore reasonable.
    \item[Primary evolved first] All of the binaries in our sample appear to be
    single-lined (from visual inspection of the \apogee\ spectra).
    Therefore, if the companion evolved first, it would now be a stellar
    remnant, and its evolution as a giant would have contributed significantly
    to the orbital circularization.
    Only 4 of the systems with measured masses have minimum mass ratios above 1,
    indicating a possible neutron star or black hole companion.
    However, an unknown fraction could have white dwarf companions.
    Still, the majority of systems in the subset with measured masses are
    consistent with low-mass main sequence companions.
    \item[First ascent] As mentioned above, only 7 of the primary stars in our
    sample are likely red clump stars (\citealt{Ting:2018}), and these systems
    all have periods between $120$--$1000~\textrm{day}$.
    This is possibly because shorter period companions would have been
    common-envelope when the evolved star reached the tip of the red giant
    branch: During this phase, rapid orbit evolution with timescales of months
    to years will cause engulfment and possible destruction of the companion
    (e.g., \citealt{Nordhaus:2010}).
    We therefore consider this to be a reasonable assumption.
\end{description}

%
%

\subsection{Exceptional systems}

There are 10 systems with $e > 0.1$ and $P/\Psurf < 6$ apparent in \figurename
s~\ref{fig:PeK} and \ref{fig:dlne}.
These binaries have short periods and large eccentricities that appear to be
discrepant with our conclusions made about tidal circularization.
Of these 10 systems, 6 have the warning flag \texttt{SUSPECT\_BROAD\_LINES} from
the \apogee\ pipeline, which suggests that these primary stars are either fast
rotators or may contain blended light from the companion star; The inferred
eccentricities for these systems may therefore be biased.
In total, 39 systems distributed over the full range of periods have this
warning flag.
The remaining 4 systems are listed in \tablename~\ref{tbl:except} and warrant
further study to understand why they have not circularized.
One possibility is that these are actually triple systems with misaligned,
long-period companions: The outer body could drive the eccentricity of the inner
companion through Kozai-Lidov oscillations (\citealt{Kozai:1962, Lidov:1962}).
Verifying this would require long-term radial velocity monitoring of these
systems.

\begin{table}[h]
    \centering
    \begin{tabular}{ c c c c c c c }
    \hline
    \texttt{APOGEE\_ID} & \logg\ & $P$ [day] & $\sigma_P$ [day] & $e$ & $\sigma_e$ & $P/\Psurf$ \\
    \hline
    2M05272385+1246432 & 3.55 & 1.0529 & 0.0001 & 0.15 & 0.02 & 1.7\\
    2M11482364+3504215 & 3.47 & 3.69   & 0.01   & 0.32 & 0.09 & 3.6\\
    2M15042105+2700026 & 2.41 & 36.68  & 0.04   & 0.61 & 0.04 & 2.5\\
    2M16503620-0054432 & 3.52 & 3.9823 & 0.0003 & 0.38 & 0.05 & 3.8\\
    \hline
    \end{tabular}
    \caption{Four systems that have small ratios $P/\Psurf$ but still have
    significant eccentricities.
    Stellar parameters are for the primary (observed) member of the systems.
    In all cases, the reported uncertainties on the surface gravities are small
    $\sigma_{\logg} < 0.1$.
    }
    \label{tbl:except}
\end{table}

Another interesting set of systems have negligible measured eccentricities with
periods between $10 < P < 100~\textrm{day}$ (see top middle and right panels of
\figurename~\ref{fig:P-e-grid}).
Analogs of these systems are also seen in the simulated population in the middle
panel of \figurename~\ref{fig:simulated}: In the simulated population, these are
simply systems that started with lower eccentricities and thus can evolve faster
to $e \approx 0$.
In the observed systems, we expect the ``spur'' of low-eccentricity systems at
periods longer than the circularization period to be a mix of systems that
started with lower eccentricity, systems in which the companion has already
evolved, and a minority of red clump stars that circularized when the primary
was at the tip of the giant branch.

\section{Conclusions}

We selected binary star systems with well-determined orbital properties from a
catalog of binaries identified using repeat radial velocity measurements from
the \apogee\ survey (\citealt{Price-Whelan:2018}).
Because of the selection function for \apogee, the majority of these systems
contain an evolved primary star that dominates the luminosity of the system, and
thus these systems are predominantly single-lined spectroscopic binaries.
The systems have a range of orbital periods and the primaries have a range of
surface gravities, indicating that they are a mix of subgiant, giant branch, and
red clump stars (\figurename~\ref{fig:apogee}).

As has been seen in many other samples of binary star systems, we find that the
short period systems have small or zero eccentricities, but above a
characteristic circularization period the systems have a range of
eccentricities; This circularization period depends on the stellar parameters of
the primary (evolved) star in the system (\figurename~\ref{fig:P-e-grid}).
If we normalize the orbital periods by the orbital period of the system with a
hypothetical companion whose orbit just grazes the surface of the primary star
(i.e. to remove the dependence on the primary star size), we find a steep
transition from eccentric to circular orbits that occurs around $P / \Psurf
\approx 10$.
This dimensionless circularization period is consistent with theoretical
predictions (\figurename~\ref{fig:simulated}):
We find that the eccentricities and observed circularization periods of binary
star systems with evolved members can be explained using standard tidal
circularization theory for stars with significant convective envelopes
(\citealt{Zahn:1977, Zahn:1989, Verbunt:1995}).

The \apogee\ survey was not designed to do binary star science
(\citealt{Majewski:2017}), but has enabled a number of stellar companion studies
because it returns to fields as a part of its survey strategy
(\citealt{Troup:2016, Badenes:2018, Price-Whelan:2018}).
Future data releases from \apogee\ will provide more visits for current stars
and nearly twice as many sources, which will allow more detailed studies of
tidal circularization and other binary star phenomena.
However, for bright stars, the end-of-mission data release of the \project{Gaia}
mission will revolutionize binary star science by providing time-series radial
velocity information for nearly 100 million stars.

\acknowledgements

It is a pleasure to thank
Matteo Cantiello (Flatiron),
David W. Hogg (NYU/Flatiron/MPIA),
Hans-Walter Rix (MPIA),
and Joshua Winn (Princeton).

The authors are pleased to acknowledge that the work reported on in this
paper was substantially performed at the TIGRESS high performance computer
center at Princeton University which is jointly supported by the Princeton
Institute for Computational Science and Engineering and the Princeton
University Office of Information Technology's Research Computing department.

\software{
    \package{Astropy} (\citealt{Astropy-Collaboration:2013}),
    \package{IPython} (\citealt{Perez:2007}),
    \package{matplotlib} (\citealt{Hunter:2007}),
    \package{numpy} (\citealt{Van-der-Walt:2011}),
    \package{scipy} (\url{https://www.scipy.org/}).
}

\facility{\sdssiv, \apogee}

\bibliographystyle{aasjournal}
\bibliography{refs}

\begin{thebibliography}{}
\expandafter\ifx\csname natexlab\endcsname\relax\def\natexlab#1{#1}\fi
\providecommand{\url}[1]{\href{#1}{#1}}
\providecommand{\dodoi}[1]{doi:~\href{http://doi.org/#1}{\nolinkurl{#1}}}
\providecommand{\doeprint}[1]{\href{http://ascl.net/#1}{\nolinkurl{http://ascl.net/#1}}}
\providecommand{\doarXiv}[1]{\href{https://arxiv.org/abs/#1}{\nolinkurl{https://arxiv.org/abs/#1}}}

\bibitem[{{Abolfathi} {et~al.}(2017){Abolfathi}, {Aguado}, {Aguilar}, {Allende
  Prieto}, {Almeida}, {Tasnim Ananna}, {Anders}, {Anderson}, {Andrews},
  {Anguiano}, \& et~al.}]{Abolfathi:2017}
{Abolfathi}, B., {Aguado}, D.~S., {Aguilar}, G., {et~al.} 2017, ArXiv e-prints.
\newblock \doarXiv{1707.09322}

\bibitem[{{Astropy Collaboration} {et~al.}(2013){Astropy Collaboration},
  {Robitaille}, {Tollerud}, {Greenfield}, {Droettboom}, {Bray}, {Aldcroft},
  {Davis}, {Ginsburg}, {Price-Whelan}, {Kerzendorf}, {Conley}, {Crighton},
  {Barbary}, {Muna}, {Ferguson}, {Grollier}, {Parikh}, {Nair}, {Unther},
  {Deil}, {Woillez}, {Conseil}, {Kramer}, {Turner}, {Singer}, {Fox}, {Weaver},
  {Zabalza}, {Edwards}, {Azalee Bostroem}, {Burke}, {Casey}, {Crawford},
  {Dencheva}, {Ely}, {Jenness}, {Labrie}, {Lim}, {Pierfederici}, {Pontzen},
  {Ptak}, {Refsdal}, {Servillat}, \& {Streicher}}]{Astropy-Collaboration:2013}
{Astropy Collaboration}, {Robitaille}, T.~P., {Tollerud}, E.~J., {et~al.} 2013,
  \aap, 558, A33, \dodoi{10.1051/0004-6361/201322068}

\bibitem[{{Badenes} {et~al.}(2018){Badenes}, {Mazzola}, {Thompson}, {Covey},
  {Freeman}, {Walker}, {Moe}, {Troup}, {Nidever}, {Allende Prieto}, {Andrews},
  {Barb{\'a}}, {Beers}, {Bovy}, {Carlberg}, {De Lee}, {Johnson}, {Lewis},
  {Majewski}, {Pinsonneault}, {Sobeck}, {Stassun}, {Stringfellow}, \&
  {Zasowski}}]{Badenes:2018}
{Badenes}, C., {Mazzola}, C., {Thompson}, T.~A., {et~al.} 2018, \apj, 854, 147,
  \dodoi{10.3847/1538-4357/aaa765}

\bibitem[{{Blanton} {et~al.}(2017){Blanton}, {Bershady}, {Abolfathi},
  {Albareti}, {Allende Prieto}, {Almeida}, {Alonso-Garc{\'{\i}}a}, {Anders},
  {Anderson}, {Andrews}, \& et~al.}]{Blanton:2017}
{Blanton}, M.~R., {Bershady}, M.~A., {Abolfathi}, B., {et~al.} 2017, \aj, 154,
  28, \dodoi{10.3847/1538-3881/aa7567}

\bibitem[{{Bluhm} {et~al.}(2016){Bluhm}, {Jones}, {Vanzi}, {Soto}, {Vos},
  {Wittenmyer}, {Drass}, {Jenkins}, {Olivares}, {Mennickent}, {Vu{\v
  c}kovi{\'c}}, {Rojo}, \& {Melo}}]{Bluhm:2016}
{Bluhm}, P., {Jones}, M.~I., {Vanzi}, L., {et~al.} 2016, \aap, 593, A133,
  \dodoi{10.1051/0004-6361/201628459}

\bibitem[{{Duch{\^e}ne} \& {Kraus}(2013)}]{Duchene:2013}
{Duch{\^e}ne}, G., \& {Kraus}, A. 2013, \araa, 51, 269,
  \dodoi{10.1146/annurev-astro-081710-102602}

\bibitem[{{Garc{\'{\i}}a P{\'e}rez} {et~al.}(2016){Garc{\'{\i}}a P{\'e}rez},
  {Allende Prieto}, {Holtzman}, {Shetrone}, {M{\'e}sz{\'a}ros}, {Bizyaev},
  {Carrera}, {Cunha}, {Garc{\'{\i}}a-Hern{\'a}ndez}, {Johnson}, {Majewski},
  {Nidever}, {Schiavon}, {Shane}, {Smith}, {Sobeck}, {Troup}, {Zamora},
  {Weinberg}, {Bovy}, {Eisenstein}, {Feuillet}, {Frinchaboy}, {Hayden},
  {Hearty}, {Nguyen}, {O'Connell}, {Pinsonneault}, {Wilson}, \&
  {Zasowski}}]{Garcia-Perez:2016}
{Garc{\'{\i}}a P{\'e}rez}, A.~E., {Allende Prieto}, C., {Holtzman}, J.~A.,
  {et~al.} 2016, \aj, 151, 144, \dodoi{10.3847/0004-6256/151/6/144}

\bibitem[{{Goodman} \& {Dickson}(1998)}]{Goodman:1998}
{Goodman}, J., \& {Dickson}, E.~S. 1998, \apj, 507, 938, \dodoi{10.1086/306348}

\bibitem[{{Gunn} {et~al.}(2006){Gunn}, {Siegmund}, {Mannery}, {Owen}, {Hull},
  {Leger}, {Carey}, {Knapp}, {York}, {Boroski}, {Kent}, {Lupton}, {Rockosi},
  {Evans}, {Waddell}, {Anderson}, {Annis}, {Barentine}, {Bartoszek}, {Bastian},
  {Bracker}, {Brewington}, {Briegel}, {Brinkmann}, {Brown}, {Carr},
  {Czarapata}, {Drennan}, {Dombeck}, {Federwitz}, {Gillespie}, {Gonzales},
  {Hansen}, {Harvanek}, {Hayes}, {Jordan}, {Kinney}, {Klaene}, {Kleinman},
  {Kron}, {Kresinski}, {Lee}, {Limmongkol}, {Lindenmeyer}, {Long}, {Loomis},
  {McGehee}, {Mantsch}, {Neilsen}, {Neswold}, {Newman}, {Nitta}, {Peoples},
  {Pier}, {Prieto}, {Prosapio}, {Rivetta}, {Schneider}, {Snedden}, \&
  {Wang}}]{Gunn:2006}
{Gunn}, J.~E., {Siegmund}, W.~A., {Mannery}, E.~J., {et~al.} 2006, \aj, 131,
  2332, \dodoi{10.1086/500975}

\bibitem[{Hunter(2007)}]{Hunter:2007}
Hunter, J.~D. 2007, Computing In Science \& Engineering, 9, 90

\bibitem[{{Hut}(1981)}]{Hut:1981}
{Hut}, P. 1981, \aap, 99, 126

\bibitem[{{Jeans}(1919)}]{Jeans:1919}
{Jeans}, J.~H. 1919, \mnras, 79, 408, \dodoi{10.1093/mnras/79.6.408}

\bibitem[{{Kjurkchieva} {et~al.}(2017){Kjurkchieva}, {Vasileva}, \&
  {Atanasova}}]{Kjurkchieva:2017}
{Kjurkchieva}, D., {Vasileva}, D., \& {Atanasova}, T. 2017, \aj, 154, 105,
  \dodoi{10.3847/1538-3881/aa83b3}

\bibitem[{{Kozai}(1962)}]{Kozai:1962}
{Kozai}, Y. 1962, \aj, 67, 591, \dodoi{10.1086/108790}

\bibitem[{{Latham} {et~al.}(2002){Latham}, {Stefanik}, {Torres}, {Davis},
  {Mazeh}, {Carney}, {Laird}, \& {Morse}}]{Latham:2002}
{Latham}, D.~W., {Stefanik}, R.~P., {Torres}, G., {et~al.} 2002, \aj, 124,
  1144, \dodoi{10.1086/341384}

\bibitem[{{Lidov}(1962)}]{Lidov:1962}
{Lidov}, M.~L. 1962, Planetary and Space Science, 9, 719,
  \dodoi{10.1016/0032-0633(62)90129-0}

\bibitem[{{Majewski} {et~al.}(2017){Majewski}, {Schiavon}, {Frinchaboy},
  {Allende Prieto}, {Barkhouser}, {Bizyaev}, {Blank}, {Brunner}, {Burton},
  {Carrera}, {Chojnowski}, {Cunha}, {Epstein}, {Fitzgerald}, {Garc{\'{\i}}a
  P{\'e}rez}, {Hearty}, {Henderson}, {Holtzman}, {Johnson}, {Lam}, {Lawler},
  {Maseman}, {M{\'e}sz{\'a}ros}, {Nelson}, {Nguyen}, {Nidever}, {Pinsonneault},
  {Shetrone}, {Smee}, {Smith}, {Stolberg}, {Skrutskie}, {Walker}, {Wilson},
  {Zasowski}, {Anders}, {Basu}, {Beland}, {Blanton}, {Bovy}, {Brownstein},
  {Carlberg}, {Chaplin}, {Chiappini}, {Eisenstein}, {Elsworth}, {Feuillet},
  {Fleming}, {Galbraith-Frew}, {Garc{\'{\i}}a}, {Garc{\'{\i}}a-Hern{\'a}ndez},
  {Gillespie}, {Girardi}, {Gunn}, {Hasselquist}, {Hayden}, {Hekker}, {Ivans},
  {Kinemuchi}, {Klaene}, {Mahadevan}, {Mathur}, {Mosser}, {Muna}, {Munn},
  {Nichol}, {O'Connell}, {Parejko}, {Robin}, {Rocha-Pinto}, {Schultheis},
  {Serenelli}, {Shane}, {Silva Aguirre}, {Sobeck}, {Thompson}, {Troup},
  {Weinberg}, \& {Zamora}}]{Majewski:2017}
{Majewski}, S.~R., {Schiavon}, R.~P., {Frinchaboy}, P.~M., {et~al.} 2017, \aj,
  154, 94, \dodoi{10.3847/1538-3881/aa784d}

\bibitem[{{Mathieu}(2005)}]{Mathieu:2005}
{Mathieu}, R.~D. 2005, in Astronomical Society of the Pacific Conference
  Series, Vol. 333, Tidal Evolution and Oscillations in Binary Stars, ed.
  A.~{Claret}, A.~{Gim{\'e}nez}, \& J.-P. {Zahn}, 26

\bibitem[{{Mathieu} \& {Mazeh}(1988)}]{Mathieu:1988}
{Mathieu}, R.~D., \& {Mazeh}, T. 1988, \apj, 326, 256, \dodoi{10.1086/166087}

\bibitem[{{Mayor} \& {Mermilliod}(1984)}]{Mayor:1984}
{Mayor}, M., \& {Mermilliod}, J.~C. 1984, in IAU Symposium, Vol. 105,
  Observational Tests of the Stellar Evolution Theory, ed. A.~{Maeder} \&
  A.~{Renzini}, 411

\bibitem[{Mazeh(2007)}]{Mazeh:2007hp}
Mazeh, T. 2007, arXiv.org

\bibitem[{{Meibom} \& {Mathieu}(2005)}]{Meibom:2005}
{Meibom}, S., \& {Mathieu}, R.~D. 2005, \apj, 620, 970, \dodoi{10.1086/427082}

\bibitem[{{Meibom} {et~al.}(2006){Meibom}, {Mathieu}, \&
  {Stassun}}]{Meibom:2006}
{Meibom}, S., {Mathieu}, R.~D., \& {Stassun}, K.~G. 2006, \apj, 653, 621,
  \dodoi{10.1086/508252}

\bibitem[{{Ness} {et~al.}(2015){Ness}, {Hogg}, {Rix}, {Ho}, \&
  {Zasowski}}]{Ness:2015}
{Ness}, M., {Hogg}, D.~W., {Rix}, H.-W., {Ho}, A.~Y.~Q., \& {Zasowski}, G.
  2015, \apj, 808, 16, \dodoi{10.1088/0004-637X/808/1/16}

\bibitem[{{Nordhaus} {et~al.}(2010){Nordhaus}, {Spiegel}, {Ibgui}, {Goodman},
  \& {Burrows}}]{Nordhaus:2010}
{Nordhaus}, J., {Spiegel}, D.~S., {Ibgui}, L., {Goodman}, J., \& {Burrows}, A.
  2010, \mnras, 408, 631, \dodoi{10.1111/j.1365-2966.2010.17155.x}

\bibitem[{{Paxton} {et~al.}(2011){Paxton}, {Bildsten}, {Dotter}, {Herwig},
  {Lesaffre}, \& {Timmes}}]{Paxton:2011}
{Paxton}, B., {Bildsten}, L., {Dotter}, A., {et~al.} 2011, \apjs, 192, 3,
  \dodoi{10.1088/0067-0049/192/1/3}

\bibitem[{P\'erez \& Granger(2007)}]{Perez:2007}
P\'erez, F., \& Granger, B.~E. 2007, Computing in Science and Engineering, 9,
  21, \dodoi{10.1109/MCSE.2007.53}

\bibitem[{{Price-Whelan} {et~al.}(2017){Price-Whelan}, {Hogg},
  {Foreman-Mackey}, \& {Rix}}]{Price-Whelan:2017}
{Price-Whelan}, A.~M., {Hogg}, D.~W., {Foreman-Mackey}, D., \& {Rix}, H.-W.
  2017, \apj, 837, 20, \dodoi{10.3847/1538-4357/aa5e50}

\bibitem[{{Price-Whelan} {et~al.}(2018){Price-Whelan}, {Hogg}, \&
  {Rix}}]{Price-Whelan:2018}
{Price-Whelan}, A.~M., {Hogg}, D.~W., \& {Rix}, H.-W. e.~a. 2018, AAS journals,
  submitted

\bibitem[{{Shporer} {et~al.}(2016){Shporer}, {Fuller}, {Isaacson}, {Hambleton},
  {Thompson}, {Pr{\v{s}}a}, {Kurtz}, {Howard}, \& {O'Leary}}]{Shporer:2016}
{Shporer}, A., {Fuller}, J., {Isaacson}, H., {et~al.} 2016, \apj, 829,
  \dodoi{10.3847/0004-637X/829/1/34}

\bibitem[{{Ting} {et~al.}(2018){Ting}, {Hawkins}, \& {Rix}}]{Ting:2018}
{Ting}, Y.-S., {Hawkins}, K., \& {Rix}, H.-W. 2018, ArXiv e-prints.
\newblock \doarXiv{1803.06650}

\bibitem[{{Troup} {et~al.}(2016){Troup}, {Nidever}, {De Lee}, {Carlberg},
  {Majewski}, {Fernandez}, {Covey}, {Chojnowski}, {Pepper}, {Nguyen},
  {Stassun}, {Nguyen}, {Wisniewski}, {Fleming}, {Bizyaev}, {Frinchaboy},
  {Garc{\'{\i}}a-Hern{\'a}ndez}, {Ge}, {Hearty}, {Meszaros}, {Pan}, {Allende
  Prieto}, {Schneider}, {Shetrone}, {Skrutskie}, {Wilson}, \&
  {Zamora}}]{Troup:2016}
{Troup}, N.~W., {Nidever}, D.~L., {De Lee}, N., {et~al.} 2016, \aj, 151, 85,
  \dodoi{10.3847/0004-6256/151/3/85}

\bibitem[{{Van der Walt} {et~al.}(2011){Van der Walt}, Colbert, \&
  Varoquaux}]{Van-der-Walt:2011}
{Van der Walt}, S., Colbert, S.~C., \& Varoquaux, G. 2011, {Computing in
  Science \& Engineering}, 13, 22,
  \dodoi{http://dx.doi.org/10.1109/MCSE.2011.37}

\bibitem[{{Verbunt} \& {Phinney}(1995)}]{Verbunt:1995}
{Verbunt}, F., \& {Phinney}, E.~S. 1995, \aap, 296, 709

\bibitem[{{Zahn}(1977)}]{Zahn:1977}
{Zahn}, J.-P. 1977, \aap, 57, 383

\bibitem[{{Zahn}(1989)}]{Zahn:1989}
---. 1989, \aap, 220, 112

\bibitem[{{Zahn}(2008)}]{Zahn:2008}
{Zahn}, J.-P. 2008, in EAS Publications Series, Vol.~29, EAS Publications
  Series, ed. M.-J. {Goupil} \& J.-P. {Zahn}, 67--90

\end{thebibliography}

\end{document}